\documentstyle[12pt,epsf]{mrsproc}
\begin{document}
\newcommand{\vareps}{\varepsilon}  
\newcommand{\De}{$\Delta$}
\newcommand{\mc}{\multicolumn}
\begin{center}
{\bf  OFFSETS AND POLARIZATION AT STRAINED AlN/GaN \\
POLAR INTERFACES
}
\end{center}

\vspace{-0.1cm}
Fabio Bernardini,$^*$ Vincenzo Fiorentini$^*$,
and David Vanderbilt$^{**}$ \\

\noindent
$*$\,{\it INFM -- Dipartimento di Scienze Fisiche, Universit\`a di 
Cagliari, I-09124 Cagliari, Italy\\
$**$\, Department of Physics and Astronomy, Rutgers University, 
Piscataway, NJ, U.S.A.}\\

\vspace{-0.6cm}
                       \begin{abstract}
\vspace{-0.4cm}
The strain induced by  lattice mismatch at the interface 
is responsible for the different value of the band discontinuities 
observed recently for the AlN/GaN (AlN on GaN) and the 
GaN/AlN (GaN on AlN) polar (0001) interface. 
We present a first-principles calculation of
valence band offsets,
interface dipoles, strain-induced piezoelectric fields,
relaxed geometric structure, and formation energies.
Our results confirm the existence of a large forward-backward asymmetry 
for this interface. 
\end{abstract}

\vspace{-0.5cm}\section{INTRODUCTION}\label{sec:intro}

\vspace{-0.1cm}
A reliable determination of the valence-band offset (VBO) at the
(0001) polar interface between wurtzite AlN and GaN is still missing.
The  few experimental investigations 
available  \cite{Waldrop.APL,Martin.APL68} 
 are in mutual disagreement, and theoretical studies
 refer either to zincblende interfaces \cite{Albanesi.TechB},
or artificially lattice-matched wurtzite  interfaces \cite{Wei}.
The latter approximation leads to a less accurate determination
for the VBO, and cannot pick up the possible forward-backward asymmetry
characteristic of lattice-mismatched interfaces.
In the case of the AlN/GaN interface, lattice mismatch amounts to 
2.5 \%, and may cause a very large asymmetry. This asymmetry has 
not yet been clearly determined experimentally (it
was not even found in early experimental work
 \cite{Martin.APL65}), being 
 hidden by the large uncertainties in the measured data.
Even the best experimental investigations available face two
kinds of problem: (i) the determination of the 
 core level alignment with the valence-band maximum
(VBM) is obtained indirectly using theoretical estimates of the 
VBM position, which (as underlined by Vogel {\it et al.} \cite{Vogel}) is affected by large systematic errors;
(ii) the existence of strong polarization fields 
in both the substrate and the overlayer tends to modify the apparent
value of the VBO deduced from the core-level shift measurements.

The present {\it ab-initio} investigation includes  all strain and
relaxation-induced effects, and  overcomes the difficulty 
in VBO determination due to polarization fields by the use
of a novel charge-decomposition technique. An estimate 
of the formation energy of the interfaces studied
is also given.

\vspace{-0.5cm}\section{BULK PROPERTIES}

\vspace{-0.1cm}
Valence-band offset calculations at lattice-mismatched interfaces
require the evaluation of the band structure  energies for  the 
bulk crystals in equilibrium, and subjected to biaxial strain. 
The calculations are done using density-functional theory in
the local-density approximation (LDA) to describe the exchange-correlation
energy, and ultrasoft pseudopotentials \cite{Vanderbilt.PRB41} 
for the electron-ion interaction. Plane-wave basis sets up to 25 Ry,
and 24 special {\bf k}-points
are found to  give fully converged values for the bulk properties. 
Since the  properties of GaN are affected by Ga 3{\it d}
states \cite{Fiorentini.PRB}, 
our Ga pseudopotential includes 3{\it d} electrons in the valence.
This yields  very good structural bulk 
parameters (see below).

In wurtzite crystals, the determination of the atomic structure 
at a given lattice constant {\it a} 
implies the calculation of the {\it c/a} and {\it u} parameters.
The equilibrium $c$ as been determined by fitting with a polynomial 
the total energy computed for six different values of {\it c},
with {\it u}  being determined for each value of {\it c/a} via
minimization of the Hellmann-Feynman forces, with a threshold of
10$^{-4}$ Hartree/bohr.
The calculated structural parameters are given in Table \ref{tab:struct}.
The structural parameters of 
AlN and GaN behave similarly under strain [$\Delta 
({\it c/a})/({\it c/a}) \sim 7 \%$,
$\Delta {\it u}/{\it u} \sim 2 \%$] with similar total 
energy variations. Instead, the effect of strain on 
the valence-band edge is very different.
A rationale for this difference is that  the 
AlN (GaN) band edge is a singlet (doublet)  
formed by the hybridization along the $c$-axis (in the $a$-plane) 
of N 2{\it s} orbitals with Al {\it p}$_z$  (Ga {\it p$_{xy}$})
states,  so that biaxial compression pushes the edge upward in GaN and
downward  in AlN.
\begin{table}[t]
\begin{center}
\caption{Predicted  structural parameters and valence band maxima
for equilibrium and
strained AlN and GaN.}

\vspace{0.2cm}
\begin{tabular}{|l|c|c|c|c|}\hline
Material   &    AlN      &   AlN      &   GaN      &  GaN        \\\hline
Substrate  &    ---      &   GaN      &   ---      &  AlN        \\\hline
{\it a}    &   5.814     &   6.04     &  6.04      &  5.814      \\\hline
{\it c/a}  &   1.619     &   1.51     &  1.6336    &  1.73       \\\hline
{\it u}    &   0.38      &   0.3927   &  0.3761    &  0.3653     \\\hline
$E_{\rm strain}$ (eV)
           &             & +0.179     &            & +0.155      \\\hline
$E_{\rm VBM}$ (eV) 
	   &--0.16       &  0.09      &--4.90      &--4.69       \\\hline
           \hline
\end{tabular}
\label{tab:struct}
\end{center}
\vspace{-0.5cm}
\end{table}

\vspace{-0.5cm}\section{BAND OFFSET}

\vspace{-0.1cm}
As pointed out by Baldereschi {\it et al.} \cite{Baldereschi.PRL}, the
valence-band offset \De E$_v$ may be split in two terms:
\[
\Delta E_v = \Delta E_{\rm VBM} + \Delta V_{el}.
\]
The first contribution $\Delta E_{\rm VBM}$ 
is the difference between the valence-band edge energy
in the two bulk materials, each edge being referred to 
the average bulk electrostatic potential.
The second contribution, the potential lineup $\Delta V_{el}$, 
is the drop of the macroscopic average of the
electrostatic potential across the interface.
The latter term requires a selfconsistent calculation of the 
electronic density distribution for the real interface system. 
Our interface has been modeled using a (GaN)$_4$/(AlN)$_4$(0001)
superlattice (see below), both 
ideal and fully relaxed. The 
material being grown epitaxially on the chosen substrate, has been
pre-strained to  have the same $a$ lattice constant as the substrate.

\begin{figure}[th]
\unitlength=1cm
\begin{center}
\begin{picture}(9,9.3)
\put(-2.0,-1.0){\epsfysize=11cm
\epsffile{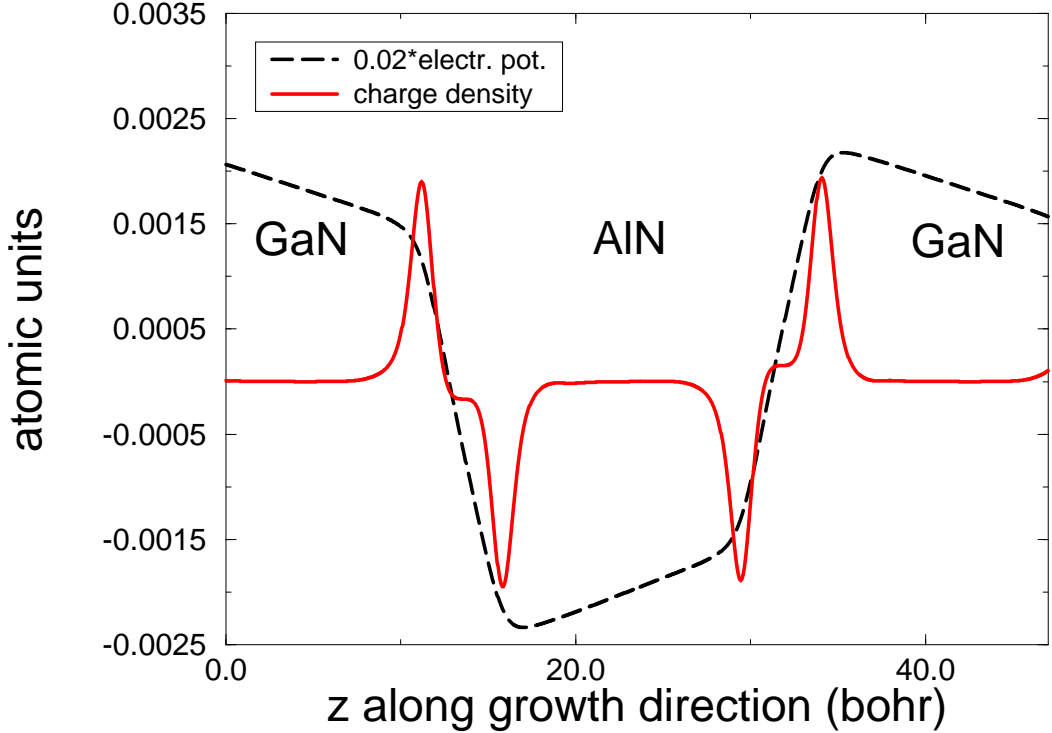}}
\end{picture}
\end{center}
\caption{Supercell electron density and electrostatic potential.
Electron density has been compensated in the two bulks by
a distribution of gaussians placed at the atomic sites.}
\label{fig:scf}
\end{figure}

\begin{figure}[th]
\unitlength=1cm
\begin{center}
\begin{picture}(9,9.3)
\put(-2.0,-1.0){\epsfysize=11cm
\epsffile{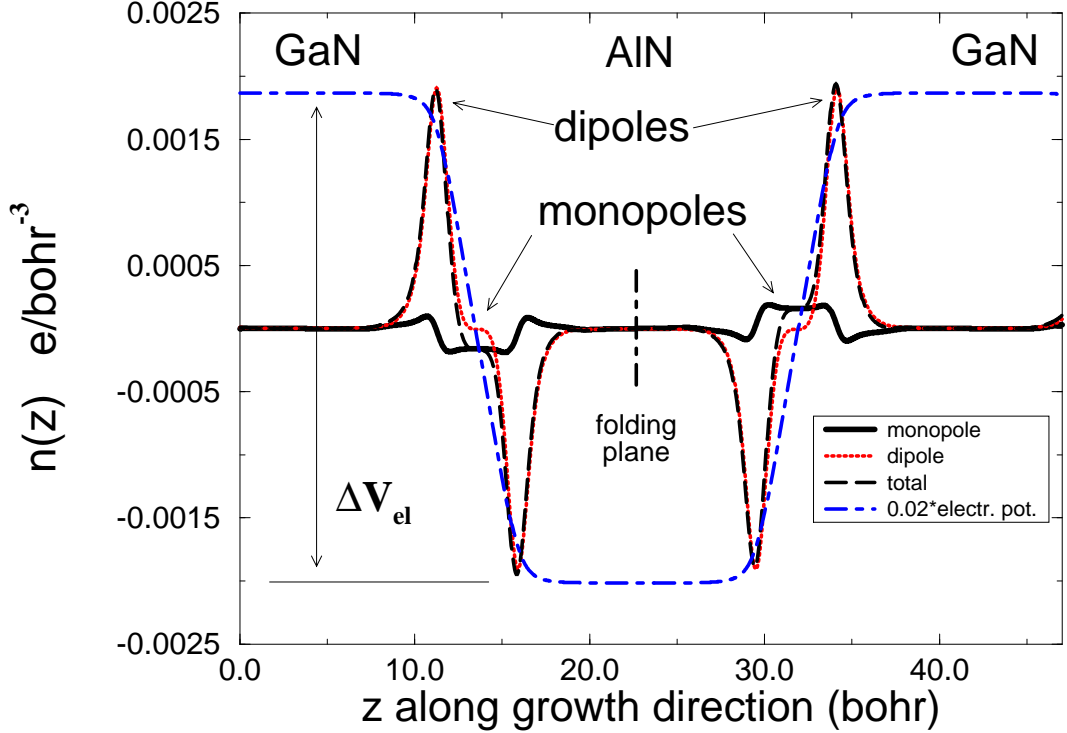}}
\end{picture}
\end{center}
\caption{Decomposition of the macroscopic average of the electronic
density (dotted line) into monopole (solid) and dipole (dashed) terms.
Such a decomposition allows the determination of the lineup term 
\De V$_{el}$ from the solution of the Poisson equation (dot-dashed) 
of the dipole term.}
\label{fig:decomp}
\end{figure}

The lineup term is customarily obtained by solving the Poisson equation
for the macroscopic average of the charge density, neutralized by a 
suitable  distribution of gaussian charges centered on the ion sites.
The potential drop across the interface is usually calculated 
as  the difference of potential values in bulk-like regions 
inside the two interfaced materials. 
This turns out to be non-trivial
for a system such as the present one,  in which the existence of 
polarization fields in the equilibrium bulk makes it 
impossible to define asymptotic values for the electrostatic 
potentials. The existence of such fields, moreover, limits the maximum 
length of our slab. 
Indeed, beyond a certain critical thickness the drop of the 
potential inside each slab would make the system metallic,
with a related transfer of charge, which
would spoil the exact determination of the lineup term.   
Our choice of  a 16-atom supercell is a compromise between the 
need to have an insulating system, and at the same time to avoid
a spurious coupling of  the interfaces at the sides of the slabs.
Tests were performed in supercells of up to 40 atoms.
In Fig. \ref{fig:scf}, we show the macroscopic average
of the charge density and of the electrostatic potential.
The potential drop is inextricably linked to the 
polarization fields within the AlN and GaN bulks.
We have circumvented this problem by employing a new method.
The basic idea is that at the polar AlN/GaN interfaces,
the existence of polarization fields reveals itself by an accumulation
of charge in the form of a {\it monopole} distribution
whose density is proportional to the difference between the
polarizations inside the two interfaced bulks.
On top of this monopole term, we have the traditional 
{\it dipole} term representing the local charge transfer across the
interface.
This dipole term is the quantity we are interested in,
as the band offset is by definition related to the dipolar part of
the potential drop. Since the monopole contributions are related to the
polarization fields, 
they must be equal and opposite for the two (geometrically inequivalent)
interfaces in our AlN/GaN superlattice.
To filter out the monopole term we superimpose the two interface
distributions by folding them around a plane placed  halfway between
the two junctions. We define the dipole term $\bar{\bar{n}}^{dip}$  as
the average of the superimposed charges, 
\[
\bar{\bar{n}}^{dip}(z-z_0) = \frac{1}{2}
                         \left[\bar{\bar{n}}(z-z_0) + 
                         \bar{\bar{n}}(z_0-z) \right], 
\] 
where $z$ is a coordinate along the $c$-axis, $z_0$ the plane position 
and $\bar{\bar{n}}$ the macroscopic average for the charge density.
The monopole term $\bar{\bar{n}}^{mono}$
is just the difference between the  
dipole term and the total macroscopic charge:
\[  
\bar{\bar{n}}^{mono}(z) = \bar{\bar{n}}(z) - \bar{\bar{n}}^{dip}(z).
\]
Such a decomposition allows a determination 
of the polarization charges and dipole terms which is  nearly 
independent of the position of the folding plane. 
Fig. \ref{fig:decomp} shows the decomposition for the AlN/GaN
interface. 
The decomposition reveals the origin of the asymmetry in the total
charge distribution, 
and at the same time  it enables us to evaluate the lineup term.
In Table \ref{tab:vbo} we report the values for the 
VBO obtained via this decomposition.
\begin{table}[th]
\caption{Valence-band offset \De $E_v$, potential lineup \De $V_{el}$, 
relaxation energies $E_{rel}$, monopole charge densities $\sigma_{int}$ and 
electric fields $\vec{E}$ in the ideal 
and relaxed the AlN/GaN (0001) interface.}  
\begin{center}
\begin{tabular}{|l|c|c|c|c|c|}\hline
Interface    & \mc{2}{c|}{AlN/GaN} &  \mc{2}{c|}{GaN/AlN} & units  \\ \hline
structure    & ideal   &  relaxed  & ideal   &  relaxed&   \\ \hline
\De $E_v$    & 0.29    &  0.20     & 1.00    &  0.85   & eV  \\ \hline
\De $V_{el}$ & 5.28    &  5.18     & 5.52    &  5.36   & eV \\ \hline 
$\sigma_{int}$& 0.029  & 0.014     & 0.022   &  0.011  & C/m$^{2}$ \\ \hline
$\vec{E}$& 32.7    & 15.6      & 24.4    &  12.9   & 10$^8$ V/m \\\hline
\hline
\end{tabular}
\end{center}
\vspace{-0.5cm}
\label{tab:vbo}
\end{table}
There is a very large forward-backward asymmetry of 0.65 eV 
between AlN/GaN and GaN/AlN interface VBOs.
This
is only marginally due to the lineup term (contributing 0.18 eV),  
its main component being the band structure term (0.47 eV).
The relaxation is responsible for comparatively small deviations
of   $\sim$ 0.1 eV from the ideal-interface values.
The relaxation pattern is characterized in both cases
by a contraction of the Al-N axial interface bond ($\sim$ --0.04 a.u.)  
and an expansion of the axial Ga-N bond ($\sim$ +0.02 a.u.).

\vspace{-0.5cm}\section{POLARIZATION}

\vspace{-0.1cm}
Supercell calculations are not the only way to obtain the 
interface charge density $\sigma_{int}$. 
As shown in Ref. \cite{Vanderbilt.PRB48},
given the polarization $P_1$ and $P_2$ and
the dielectric constants $\vareps_1$ and $\vareps_2$ 
of  the component materials,
$\sigma_{int}$ is given by
\begin{equation}
\sigma_{int} = \pm ~2\, (P_2 - P_1)/(\vareps_1+\vareps_2).
\label{eq:sigma}
\end{equation}
in periodic boundary conditions. We have calculated the macroscopic
polarization for equilibrium and strained GaN and AlN via the Berry
phase technique of Ref. \cite{King.PRB47}. 
The (high-frequency) dielectric constants of AlN and GaN  have been calculated
using the relation
\[
\Delta P_{\rm T} = \vareps_{\infty} \Delta P_{\rm L}, 
\]
where $\Delta P_{\rm T}$ is the (so-called transverse) polarization 
change induced by a
small  cation sublattice displacement in the bulk
in zero field, and  $\Delta P_{\rm L}$ is the (so-called longitudinal)
 polarization change due to a uniform displacement of few cation
planes in a periodic bulk supercell. In the latter, a depolarizing
field is present due to the periodic boundary conditions.
%
%
As a by-product of our calculations, we obtained the
Born effective charges for AlN and GaN which,  as expected for highly polar 
semiconductors, are quite close to the nominal ionicity. 
The results are shown in Table \ref{tab:pol}.
\vspace{-0.3cm}
\begin{table}[bh]
\caption{Polarization in GaN and AlN:
electronic $P_{el}$ and total $P_{tot}$ polarization,
 derivative of the latter with respect to $u$, Born effective
charge, and dielectric constant are shown.}
\begin{center}
\begin{tabular}{|c|c|c|c|c|c|c|c|} \hline
System &  $a$   & $P_{el}$&$P_{tot}$&  $\partial P 
                                   / \partial u$ & $Z^*$ &$\varepsilon_{\infty}$  \\\hline
-----  &  bohr  & C/m$^2$ &C/m$^2$  & C/m$^2$ & $e$ & -----\\\hline
AlN    &  5.814 &--0.178  &--0.0812 &  10.51  & 2.69  &  4.59 \\\hline 
AlN    &  6.04  &--0.492  &--0.1712 &   9.95  & 2.74  &  4.64 \\\hline 
GaN    &  5.814 & +0.223  & +0.0343 &  10.69  & 2.74  &  5.27 \\\hline 
GaN    &  6.04  &--0.0511 &--0.0308 &   9.88  & 2.72  &  5.52 \\\hline 
\hline
\end{tabular}
\end{center}
\vspace{-0.5cm}
\label{tab:pol}
\end{table}
Substituting  $P_{tot}$ and $\vareps$ in Eq.\ref{eq:sigma}  we 
obtain for the AlN/GaN (GaN/AlN) interface a monopole density 
of 0.028 (0.023) C/m$^2$ in embarrassing agreement with the 
outcomes of the supercell calculations. This proves directly that the
sources of the internal fields in the interface system are the
 charges accumulated at the interface by the
bulk polarization effects. Also, the latter finding provides an a
posteriori justification of our somewhat ad-hoc charge decomposition
 procedure.

\vspace{-0.5cm}\section{FORMATION ENERGIES}

\vspace{-0.1cm}
An important issue for the present selfconsistent calculation is the
evaluation of the formation energy for the AlN/GaN interfaces.
Contrary to the case of non-polar interfaces, for the wurtzite
(0001) system it is impossible to build a superlattice with symmetric
interfaces. This means that only the
average value of the formation energy for the two interfaces 
can be obtained  from a total-energy calculation. 
We define the average formation energy for the AlN/GaN interface as
\[
E_f = \frac{1}{2A} \left[ E^{\rm tot} - n^{\rm Ga}\mu^{\rm GaN} 
			       - n^{\rm Al}\mu^{\rm AlN}\right],
\]
where $\mu^X$ are the total energies per pair of GaN and AlN,
$n^{X}$ the number of Ga and Al atoms, $E^{\rm tot}$ the supercell
total energy and $A$ its cross-sectional area.  
A reliable determination of $E_f$ requires equivalent
k-point sampling for  bulk and interface calculations.
This is easily accomplished if the interface is lattice-matched. 
In the present case, the supercell length is not simply an integer
multiple of the bulk unit cell of either constituent material.
This means that an exact equivalence between k-point meshes cannot
be achieved.
A good approximation for $E_f$
can however be obtained by defining, for each component material,
an auxiliary bulk cell having the same 
lattice constant {\it a}, and an axial length 
${\it\bar{c}}$  being a sub-multiple
of the supercell length ${\it l}$.
This value in the present case is just the average of ${\it c_{\rm
AlN}}$ and  
${\it c_{\it GaN}}$. It is then possible to downfold exactly the
supercell mesh  into the auxiliary bulk cell.
The next step is to uniformly scale the k-points coordinates 
to adapt the mesh to the real value of $c$.
We should point out that the accuracy of this procedure (compared with
an exact computation of the energy integral over the IBZ) increases 
with the number of points in the mesh. It is therefore
possible to find a suitable mesh to accomplish any required accuracy.
\vspace{-0.5cm}
\begin{table}[h]
\caption{Average formation energy for the AlN/GaN (0001) interfaces.}
\begin{center}
\begin{tabular}{|l|c|c|c|c|c|}\hline
Interface    & \mc{2}{c|}{AlN/GaN} &  \mc{2}{c|}{GaN/AlN} & units  \\ \hline
             & ideal   &  relaxed  & ideal   &  relaxed   &   \\ \hline
$E_f$        &--3.4    &--16.4     & +11.7 & --6.2   &meV  \\ \hline 
\hline
\end{tabular}
\end{center}
\label{tab:eform}
\end{table}
\vspace{-0.6cm}
The results for the formation energies reported in
Table \ref{tab:eform} have been obtained using
a 6-point Chadi-Cohen mesh \cite{Chadi.Cohen}
in the supercell, which when
downfolded in the auxiliary cell produces 24 special points.
We estimate the k-point sampling error in the formation energies
to be $\sim$ 10 meV.
It should be noted that such low formation energies are not 
surprising when compared with the results obtained by Chetty
{\it et al.} \cite{Chetty.PRB} for GaAs/AlAs(111) interfaces.




\vspace{-0.6cm}
\begin{thebibliography}{99}
{\small

\vspace{-0.35cm}
\bibitem{Waldrop.APL}
J. W. Waldrop and R. W. Grant, Appl. Phys. Lett. {\bf 68}, 2879 (1996).

\vspace{-0.35cm}
\bibitem{Martin.APL68}
G. Martin, A. Botchkarev, A. Rockett, and H. Morkoc, Appl. Phys. Lett.
{\bf 68}, 2541 (1996).

\vspace{-0.35cm}
\bibitem{Albanesi.TechB}
E. Albanesi, W. R. L. Lambrecht, and B. Segall, J. Vac. Sci. Tech. 
{\bf B  12}, 2470 (1994).

\vspace{-0.35cm}
\bibitem{Wei}
S. Wei and A. Zunger, to be published.

\vspace{-0.35cm}
\bibitem{Martin.APL65}
G. Martin {\it et al.}, Appl. Phys. Lett. {\bf 65}, 610 (1994).   

\vspace{-0.35cm}
\bibitem{Vogel} 
D. Vogel, P. Kr\"uger and J. Pollmann, to be published

\vspace{-0.35cm}
\bibitem{Vanderbilt.PRB41}
D. Vanderbilt, Phys. Rev. {\bf 41}, 7892 (1990).

\vspace{-0.35cm}
\bibitem{Fiorentini.PRB}
V. Fiorentini {\it et al.}, Phys. Rev. B {\bf 47},
13353 (1993); V. Fiorentini {\it et al.}, Proc. ICPS-22,
D. J. Lockwood ed. (World
Scientific, Singapore 1995), p.137; 
A. Satta  {\it et al.},
MRS Proc. {\bf 395}, 515  (1996).  

\vspace{-0.35cm}
\bibitem{Baldereschi.PRL}
A. Baldereschi, S. Baroni, and R. Resta, Phys. Rev. Lett. 
{\bf 61}, 734 (1988).

\vspace{-0.35cm}
\bibitem{Vanderbilt.PRB48}
D. Vanderbilt and R. D. King-Smith, Phys. Rev. B {\bf 48}, 4442 (1993).

\vspace{-0.35cm}
\bibitem{King.PRB47}
R. D. King-Smith and D. Vanderbilt, Phys. Rev. B {\bf 47}, 1651 (1992). 
}
\vspace{-0.35cm}
\bibitem{Chadi.Cohen}
D. J. Chadi and Marvin L. Cohen, Phys. Rev. B {\bf 8}, 5747 (1973).

\vspace{-0.35cm}
\bibitem{Chetty.PRB}
N. Chetty and R. M. Martin, Phys. Rev. B {\bf 45}, 6089 (1992).

\end{thebibliography}
\end{document}